\documentclass[journal,onecolumn,a4paper]{IEEEtran}

\IEEEoverridecommandlockouts
\usepackage{graphicx}                                      
\usepackage{amssymb,amsthm,bm}  
\usepackage[cmex10]{amsmath} 

\usepackage{multirow}   
\usepackage{tikz}                   
\newtheorem{theorem}{Theorem}

\newtheorem{remark}{Remark}
\newtheorem{definition}{Definition}

\newtheorem{proposition}{Proposition}

\newtheorem{example}{Example}

\newcommand{\utag}[2]{\mathop{#2}\limits^{\text{(#1)}}}
\newcommand{\uref}[1]{(#1)}

\long\def\symbolfootnote[#1]#2{\begingroup
\def\thefootnote{\fnsymbol{footnote}}\footnote[#1]{#2}\endgroup}

\usepackage{xcolor}
\definecolor{SkyBlue}{RGB}{240,255,255}
\definecolor{LavenderBlush}{RGB}{218,112,214}
\definecolor{BlanchedAlmond}{RGB}{255,235,205}

\newcommand{\cB}{\mathcal{B}}

\newcommand{\cN}{\mathcal{N}}

\newcommand{\cQ}{\mathcal{Q}}

\usepackage{mathrsfs}

\newcommand{\Ie}{\textit{i.e., }}

\usepackage{diagbox}
\usepackage{color}

\usepackage{tikz}
\usetikzlibrary{automata,positioning,chains,fit,shapes,calc}
\usetikzlibrary{arrows}
\usepackage{tkz-graph}
\usepackage{caption}
\usepackage{booktabs}
\usepackage{cite}
\usepackage{url}

\makeatletter
\newcommand{\mathleft}{\@fleqntrue\@mathmargin0pt}
\newcommand{\mathcenter}{\@fleqnfalse}
\makeatother
\newcommand{\Pb}[1]{\Pr\left(#1\right)}
\newcommand{\pb}[1]{p\left(#1\right)}
\newcommand{\wb}[1]{w\left(#1\right)}

\begin{document}

\title{Preserving ON-OFF Privacy for Past and Future Requests\\
\thanks{This work was supported by NSF Grant CCF 1817635.} 
}


 \author{
   \IEEEauthorblockN{Fangwei Ye, Carolina Naim, Salim El Rouayheb} \\
   \IEEEauthorblockA{Department of Electrical and Computer Engineering, Rutgers University\\
                     Emails: \{fangwei.ye, carolina.naim,  salim.elrouayheb\}@rutgers.edu
                     }
                    
}

\maketitle

\begin{abstract}
We study the ON-OFF privacy problem. At each time, the user is interested in the latest message of  one of $N$ sources. Moreover, the user  is assumed to be incentivized to turn privacy ON  or OFF whether he/she needs it or not.  When privacy is ON, the user wants to keep private which source he/she is interested in. The challenge here is that the user's behavior is correlated over time. Therefore,  the user cannot simply ignore privacy when privacy is OFF, because this may leak information about his/her behavior when privacy was ON due to correlation.

We model the user's requests  by a Markov chain. The goal  is to design ON-OFF privacy schemes with optimal download rate that ensure privacy for past and future requests. The user is assumed to know future requests within a window of positive size $\omega$ and uses it to construct privacy-preserving queries. In this paper, we construct ON-OFF privacy schemes for $N=2$ sources and prove their optimality. 

\end{abstract}



 \section{Introduction}
 \label{introduction}

 Privacy of online users has become a major concern. Without agreeing to it, users unknowingly leak valuable personal information, such as sex, age, health disorders, political views, etc., through their daily online activities. Several existing privacy-preserving solutions can be utilized to ensure a desired level of privacy for the user, such as 
   anonymity \cite{Sweeney_2002}, differential privacy \cite{Dwork_2006}, private information retrieval \cite{Chor_1995}, to name a few.


In all the privacy problems above, it is assumed that the user always wants to be private. Privacy, however, is expensive. Privacy-preserving protocols  incur higher computational  costs on the service provider, and typically lead to degraded quality of service and larger delays at the user side \cite{DSRHR}.  

This motivates us to think of privacy as an expensive utility, which should be turned OFF when not needed. Much like one turns off the lights before leaving home. The user may want to turn his/her privacy ON or OFF depending  on the internet connection he/she is using, his/her location or his/her device used to get online, etc. This behavior of the user may be incentivized  by the service providers who encourage him/her to require privacy only when it is needed. 

The challenge in designing algorithms that enable privacy to be switched between ON and OFF, and vice versa, is that the user's behavior is correlated over time. This is essentially true because the user's choices are personal and are not independent over time. For instance, a user watching online videos, will most likely pick the next video to watch from a  suggested personalized list that is specifically curated for him/her. Therefore,  the user cannot simply ignore privacy when privacy is OFF, because this may leak information about his/her behavior when privacy was ON due to correlation.


To capture this challenge, the authors introduced the ON-OFF privacy problem in \cite{Naim_2019}.
A user is interested in the latest message generated by one of $N$ sources. Think, for example, a user is subscribed to $N=2$ political YouTube channels, one is pro-right and one is pro-left. 
Occasionally, the user wants to watch the latest video on one of these channels. He/she has a choice between turning privacy ON or OFF. When privacy is ON,  the user is not interested in hiding which particular video he/she wants to watch. Rather, he/she is interested in hiding the  channel on which that video is posted, because he/she does not want to reveal his/her political interests. In general, when privacy is ON,  the user wants to hide which message of the $N$ sources he/she is interested in.





In \cite{Naim_2019}, we studied  ON-OFF privacy   in which it was required  to ensure privacy for  past requests for which privacy was turned ON. In this paper, we consider a more stringent privacy requirement and want to preserve privacy for both {\em past and future} requests. We follow a   setup similar to the one in \cite{Naim_2019} in which  the user's request are modeled by a Markov chain, but with one significant difference. We assume here that the user knows the requests in a small window of positive size $\omega>0$ in the future. In practice, this may happen in applications where the user can queue up his/her requests, such as when watching online videos.


Under this new setting, we study the download rate, which is measured by the ratio of  the average length of downloaded data to the message length. We characterize the optimal download rate for the system with $N=2$ sources and provide explicit constructions of ON-OFF privacy schemes that achieve it. One interesting implication of our result, is that the optimal rate does not depend on the window   size. Thus, a window of size $\omega = 1$ is sufficient to achieve the optimal  rate.

\section{Problem Formulation}
\label{sec:formulation}

There is a single server storing $N$ information sources indexed by $\cN:=\{1,\ldots,N\}$. Each source generates an independent message $W_{x,t}$ at time $t$, where $x~ \in~ \cN$. We assume that $t \in \mathbb{N}$ throughout this paper.

A  user is interested in one of the sources at each time, and wishes to retrieve the latest message generated by the corresponding source. In particular, let $X_t$ be the index of the desired source at time $t$, and in the sequel we call $X_t$ the user's request. By slightly abusing the notation, we denote the latest message generated by the desired source $X_t$ by $W_{X_t,t}$, and the user wishes to retrieve the message $W_{X_t,t}$. We assume that the messages $\{W_{x,t}: x \in \cN, t \in \mathbb{N}\}$ are mutually independent, and each message consists of $L$ symbols. Without loss of generality, we assume that each of the messages is uniformly distributed over $\{0,1\}^L$, \Ie 	$H\left(W_{x,t}\right) = L$, and

\begin{equation}
 	H\left(W_{x,t}: x \in \cN, t \in \mathbb{N} \right) = \sum_{x,t} H\left(W_{x,t}\right).
\end{equation}

The user's requests are generated by a discrete-time information source $\{X_t: t \geq 0\}$. 
In this paper, we are particularly interested in the case where the requests $\{X_t: t \geq 0\}$ are Markov. The transition matrix $M$ of the Markov chain is assumed to be known by both the server and the user. 

At time $t$, the user may or may not wish to keep the identity of the source being interested in. Specifically, the \emph{privacy mode} $F_t$ at time $t$ can be either ON or OFF, where $F_t$ is ON when the user wishes to keep $X_t$ private, while $F_t$ is OFF when the user is not concerned with privacy. The privacy mode is also assumed to be known by the server.

The user is allowed to generate unlimited local randomness, and we are not interested in the amount of randomness used. Therefore, we assume without loss of generality that the random variables $\{S_t : t \geq 0\}$, representing the local randomness, are mutually independent. 

All information sources are assumed to be independent, that is,  the user's requests $\{X_t: t \geq 0\}$, the privacy mode $\{F_t:t \geq 0\}$, the messages $\{W_{x,t}: x \in \cN, t \geq 0\}$ and the local randomness $\{S_t: t \geq 0 \}$ are mutually independent. 

At time $t$, the user will construct a query $Q_t$ and send it to the server. Upon receiving the query, the server responds by producing an answer $Y_t$. After receiving the answer, the user should be able to decode $W_{X_t,t}$ correctly.

We assume that the user knows the future requests in a window of positive size $\omega$. This means at time $t$, the user knows the future requests $\{X_{t+1},\ldots,X_{t+\omega}\}$
in addition to all past (including current) requests $\{X_0,\ldots,X_t\}$. In practice, it often happens that the user has some side information to predict his/her requests in the near future. Later, we will show that only a window of size $\omega=1$ is needed.

The query $Q_t$ at time $t$ is generated by the query encoding function $\phi_t$, which is
assumed to be a function of the \emph{causal} information, \Ie previous requests and local randomness $\{X_i,S_i: i \leq t \}$, and  future requests  $\{X_{t+1},\ldots,X_{t+\omega}\}$ for some $\omega \in \mathbb{N}$. 
Hence, we assume that  
\begin{equation}
	Q_t = \phi_t \left( X_{[t+\omega]},S_{[t]} \right),
\end{equation}
where $[t+\omega]:=\{0,1,\ldots,t+\omega\}$.

Accordingly, the answer $Y_t$ of the server is given by the answer encoding function $\rho_t$, which 
is assumed to be a function of the query $Q_t$ and the latest messages, \Ie 
\begin{equation}
 	Y_t = \rho_t \left(Q_t,W_{1,t},\ldots, W_{N,t}  \right).
\end{equation} 

To facilitate our discussion, we define the length function of the answer as follows.
Since the length of the answer $Y_t$ is determined by the query $Q_t$, let $\ell(Q_t)$ be the length of $Y_t$ and the average length of the answer at time $t$ is given by \begin{equation}
  \ell_{t} = \mathbb{E}_{Q_{t}} [\ell\left(Q_t\right)],
\end{equation}
where $\mathbb{E[\cdot]}$ is the
expectation operator.

The query and answer functions need to satisfy the following decodable and privacy constraints.
\begin{enumerate}
	\item Decodability: For any time $t$, the user should be able to recover the desired message from the answer with zero-error probability, \Ie
	\begin{equation}
	\label{eq:decode}
		H\left(W_{X_t,t}|Y_{t}\right) = 0, \quad \forall t\in \mathbb{N}.
	\end{equation}
	\item 
	
	Privacy: For any time $t$, the user's requests over time where the privacy is required should not be revealed to the server, \Ie
	\begin{equation}
	\label{eq:privacy}
		I \left(X_{\cB_t};Q_{[t]}\right) = 0, \quad \forall t\in \mathbb{N},
	\end{equation}
	where $\cB_t := \{i:  i \leq t, F_i=\text{ON}\}\cup \{i: i \geq t+1\}$, and $[t]: =\{0,1,\ldots,t\}$.


%
\end{enumerate}

We would like to clarify the privacy requirement in \eqref{eq:privacy}. The user does not know whether privacy is ON or OFF in the future. For this reason, we have adopted a worst-case formulation in the privacy constraint by assuming that privacy is always ON in the future.

For any message length $L$, the tuple $\left(\ell_t: t \in \mathbb{N}\right)$ is said to be achievable if there exists a code satisfying the decodability and the privacy constraint. 
The efficiency of the code can be measured by $L/\ell_t$. Hence, we define the achievable region by the convention as follows:
\begin{definition}
The rate tuple $\left(R_t:t \in \mathbb{N}\right)$ is achievable if there exists a code with message length $L$ and average download cost $\ell_t$ 
such that $R_t \leq L /\ell_t$.
\end{definition}

Before proceeding to the results, we would like to mention that coded retrieval is not helpful in this problem. The point can be formally argued by dividing the possible queries to $2^{N}$ subsets, each of which corresponds to the decodability of a subset of the latest messages. Details can be found in \cite{Naim_2019}. For this reason, we only consider that $Q_t$ takes value in $\cQ=2^{\cN}$ in the following sections.

\section{Main result}

In this section, we present the main result of this paper, that is, the characterization of the achievable region for the two-sources system, \Ie $N = 2$. For clarity, we will use $A$ and $B$ to denote the two sources, that is, each $X_t$ takes values in $\mathcal{N}=\{A,B\}$. Correspondingly, the query $Q_t$ takes values in $\mathcal{Q}=\left\{\{A\},\{B\},\{A,B\}\right\}$. We do not distinguish between $A$ and $\{A\}$ in our notation, and $\{A,B\}$ will be written as $AB$.

Before stating our main result, we need to set up some useful notations. 
For simplicity, we assume that $F_0 = \text{ON}$. For any $t$, let $F^{-}(t) := \max\{i: i \leq t, F_i = \text{ON}\}$, \Ie $F^{-}(t)$ is the latest time such that the privacy is ON.  For our analysis, it is convenient to define $U_t := \left(X_{F^{-}(t)},X_{t+1}\right)  \in \cN^2$, which represents the last request when privacy was ON and the next request of the user at time $t$. 



We will need $\pb{x_t|u_t}$, which is given by
\[
\pb{x_t|u_t} =\frac{\pb{x_{t+1}|x_t}p\left(x_t|x_{F^{-}(t)}\right)}{p\left(x_{t+1}|x_{F^{-}(t)}\right)}.
\] 
Here,  $\pb{x_{t+1}|x_t}$, $p\left(x_t|x_{F^{-}(t)}\right)$ and 
$p\left(x_{t+1}|x_{F^{-}(t)}\right)$ can be determined from $M$, $M^{t-F^{-}(t)}$ and $M^{t+1-F^{-}(t)}$ respectively, where $M$ is the transition matrix of the Markov chain representing the user's requests.
Moreover, we  introduce the following definition:
\begin{equation}
\label{eq:pi}
  \pi(x_t) :=  \min_{u_t \in \cN^2}\pb{x_t|u_t}, ~\forall x_t \in \cN.  
\end{equation}
In other words, if we write $\pb{x_t|u_t}$ as a $N^2 \times N$ probability transition matrix,  $\pi(x_t)$ is the minimum value of each column.

Now, we are ready to state the main result in the following theorem.
\begin{theorem}
\label{thm:main}
Suppose that $\{X_t: t \geq 0\}$ is a Markov process with the transition matrix $M$. The rate tuple $\left(R_t:t \in \mathbb{N}\right)$ is achievable if and only if  
  \begin{equation}
  \label{eq:rate}
      \frac{1}{R_t} \geq  2- \sum_{x_t \in \cN} \pi(x_t).
  \end{equation}
\end{theorem}

To prove the theorem, we will give an explicit scheme that achieves the rate given in the R.H.S of \eqref{eq:rate} in Section~\ref{sec:achievable} and prove its optimality in Section~\ref{sec:converse}. Before that, we give an example to illustrate the rate given in \eqref{eq:rate}.
\begin{example}
\label{example1}
	Consider $(F_0,F_1)=(\text{ON},\text{OFF})$ and the transition matrix of the Markov chain is given by 
	\[M=
	\begin{bmatrix}
       1 - \alpha &  \alpha          \\
       \alpha & 1- \alpha 
    \end{bmatrix}, ~ 0 \leq \alpha \leq \frac{1}{2},
	\]
	where $M_{i,j}$ is the transition probability from source $i$ to source $j$ (assuming source $1$ is $A$ and source $2$ is $B$). 

Consider the rate at $t=1$. From \eqref{eq:rate}, we have
\begin{equation*}
	\frac{1}{R_1} \geq 2 - \frac{2\alpha^2}{\alpha^2+(1- \alpha)^2},
\end{equation*}
which means that it is not necessary for the user to download both messages except when $\alpha = 0$.
When $\alpha=0.5$, $R_1\geq 1$. The reason is that at each time the user simply downloads only his/her desired message when the requests are independent.

\end{example}

Few remarks about the theorem are due here.
\begin{remark}
In our model, we have assumed that the user knows the future requests within a window of   positive size $\omega \geq 1$. An interesting implication of Theorem~\ref{thm:main} is that the optimal rate does not depend on the window size. This means that increasing the window size into the future beyond one does not increase the rate.
The case when the user does not know any future requests, \Ie  $\omega=0$, falls into a different model, which was studied in \cite{Naim_2019}. 

\end{remark}
\begin{remark}
If $F_t=\text{ON}$, we have $U_t=(X_t,X_{t+1})$, and then  we can easily see that $R_t$ is achievable if and only if $R_t \leq \frac{1}{2}$ from \eqref{eq:pi} and \eqref{eq:rate}, which means that it is necessary to download two messages. This is consistent with the well-known result~\cite{Chor_1995}.
\end{remark}

	





\section{Proof of Theorem~\ref{thm:main}: Achievability}
\label{sec:achievable}

\subsection{ON-OFF Privacy Scheme}
Here, we describe our query encoding function as defined in Section~\ref{sec:formulation}. The query $Q_t$ is encoded from $X_t$, $U_t$ and $S_t$, \Ie 
\[Q_t = \phi_t \left(U_t,X_t,S_t \right).\] 
Since we are not interested in the local randomness used, instead of writing $\phi_t$ explicitly, the function $\phi_t$ can be completely described by the probability distribution $\wb{q_t|x_t,u_t}$, which is given by
\begin{table}[!htpb]
\centering
\begin{tabular}{cl|ccc}
& $q_t$ & $x_t$ & $\bar{x}_t$ & $AB$ \\ 
\hline
\multicolumn{2}{c|}{\multirow{2}{*}{$w(q_t|x_t,u_t)$}} & \multirow{2}{*}{$\frac{\pi(x_t)}{p(x_t|u_t)}$} & \multirow{2}{*}{0} & \multirow{2}{*}{$1- \frac{\pi(x_t)}{p(x_t|u_t)}$} \\
\multicolumn{2}{c|}{}  &    &   &                   
\end{tabular}
\end{table}

Here, $\bar{x}_t$ is defined as $ \{A,B\}\setminus\{x_t\}$.
Since $q_t \neq \bar{x}_t$ is always true, for notational simplicity, we write the encoding function $w(q_t|x_t,u_t)$ as 
\begin{equation}
\label{eq:encoding-function}
\wb{q_t|x_t,u_t} = 
	\begin{cases}
	\frac{\pi(x_t)}{\pb{x_t|u_t}}, & |q_t|=1, \\
	1- \frac{\pi(x_t)}{\pb{x_t|u_t}}, & |q_t| = 2.
	\end{cases}
\end{equation}
\begin{example}
Let us adopt the same setting as in Example~\ref{example1}. 
Suppose that at time $t=1$, the user wants source $A$, \Ie $X_1=A$, and we need to determine the query $Q_1$. First, we determine
\begin{equation*}
    \pi\left(x_1\right) = \frac{2\alpha^2}{1+\left(1-2\alpha\right)^2}.
\end{equation*}
In our scheme in \eqref{eq:encoding-function}, $Q_1$ will be dependent on $X_0$ and $X_2$. Suppose that $X_0=X_2=A$, and then $Q_1$ will be given by
\begin{equation*}
\wb{q_1|x_1,u_1} =  
	\begin{cases}
	\frac{\alpha^2}{(1-\alpha)^2}, & |q_1|=1, \\
	\frac{1-2\alpha}{(1-\alpha)^2}, & |q_1| = 2.
	\end{cases}
\end{equation*}
In other words, if $X_0=X_1=X_2=A$, then the user will toss a biased coin such that with probability $\frac{\alpha^2}{(1-\alpha)^2}$, he/she will download only the message generated by source $A$ and with probability $\frac{1-2\alpha}{(1-\alpha)^2}$, he/she will download both messages. 

\end{example}

\subsection{Rate}

We first show that the given coding scheme achieves the rate
\[R_t = \frac{1}{2- \sum\limits_{x_t} \pi(x_t) }.\]

Since
\[\pb{q_t}  = \sum_{x_t,u_t} \pb{x_t,u_t} \wb{q_t|x_t,u_t}, \]
by substituting \eqref{eq:encoding-function}, we have 
\begin{equation}
\label{eq:temple-qt}
	\pb{q_t}  =  
	\begin{cases}
	\sum\limits_{x_t,u_t}\pb{u_t}  \pi(x_t), & |q_t|=1, \\
	1- \sum\limits_{x_t,u_t}\pb{u_t} \pi(x_t), & |q_t| = 2.
	\end{cases}
\end{equation}
Note that $\pi(x_t)$ is independent of $u_t$, so \eqref{eq:temple-qt} can be written as
\begin{equation}
\label{eq:qt}
	\pb{q_t}  =  
	\begin{cases}
	\sum\limits_{x_t}\pi(x_t), & |q_t|=1, \\
	1- \sum\limits_{x_t} \pi(x_t), & |q_t| = 2,
	\end{cases}
\end{equation}
which immediately gives that
\[\frac{1}{R_t} = \frac{\ell_t}{L} = \mathbb{E}\left[|Q_t|\right] = 2- \sum_{x_t} \pi(x_t). \]

\subsection{Privacy}
It remains to show that the encoding function given in \eqref{eq:encoding-function} satisfies the privacy constraint in \eqref{eq:privacy}.
We prove this by induction on $t$. 

First, consider the base case where $t=0$. Since $F_0 =\text{ON}$, we know that $Q_0=AB$ from \eqref{eq:encoding-function}, so we have
\[I\left(X_{\cB_0};Q_{[0]}\right) = 0.\]
Now, we start the inductive step. Assume that
\begin{equation}
\label{eq:inductive}
	I\left(X_{\cB_{t-1}};Q_{[t-1]}\right) = 0,
\end{equation}
we need to show that
\[I\left(X_{\cB_{t}};Q_{[t]}\right) = 0.\] 

Towards this end, consider
\begin{equation*}
    I\left(X_{\cB_{t}};Q_{[t]}\right) = \underbrace{I\left(X_{\cB_{t}};Q_{[t-1]}\right)}_{I_1} + I\left(X_{\cB_{t}};Q_t|Q_{[t-1]}\right),
\end{equation*}
where $I\left(X_{\cB_{t}};Q_t|Q_{[t-1]}\right)$, the second term in the summation above, can be written as
\begin{align*}
     & I\left(X_{\cB_{t}};Q_t|Q_{[t-1]}\right) \\ 
     & = ~ I\left(U_t;Q_t|Q_{[t-1]}\right) + I\left(X_{\cB_{t}}\backslash U_t;Q_t|U_t,Q_{[t-1]}\right) \\
	& = ~ I\left(U_t;Q_{[t]}\right) \hspace{-3pt}- \hspace{-2pt} \underbrace{I\left(U_t;Q_{[t-1]}\right)}_{I_2}  \hspace{-2pt}+  \underbrace{I\left(X_{\cB_{t}}\backslash U_t;Q_t|U_t,Q_{[t-1]}\right)}_{I_3}.
\end{align*}
Thus, we have
\begin{equation}
\label{eq:last-step}
    I\left(X_{\cB_{t}};Q_{[t]}\right) = I\left(U_t;Q_{[t]}\right) + I_1 - I_2 + I_3.
\end{equation}

\begin{proposition}
\label{proposition:1}
$I_1=I_2=I_3 = 0$.
\end{proposition}
  This proposition is mainly due to the causality of the encoding function and the Markovity of the user's requests. The proof details will be given at the end of this section. 


It remains to show that $I\left(U_t;Q_{[t]}\right) = 0$, which can be equivalently written as $\pb{u_t|q_{[t]}} = \pb{u_t}$. To see this, consider
\begin{align}
	\pb{u_t|q_{[t]}} 
	& =  \sum_{x_t} \pb{u_t,x_t|q_t,q_{[t-1]}} \nonumber\\
	& =  \sum_{x_t} \frac{\pb{u_t,x_t,q_t|q_{[t-1]}}}{\pb{q_t|q_{[t-1]}}} \nonumber\\
	& = \frac{ \sum_{x_t} \pb{u_t,x_t|q_{[t-1]}} \pb{q_t|u_t,x_t,q_{[t-1]}}}{\pb{q_t|q_{[t-1]}}} \nonumber\\  
	& = \frac{ \sum_{x_t} \pb{u_t,x_t|q_{[t-1]}}  \pb{q_t|u_t,x_t,q_{[t-1]}}}{\sum_{x_t,u_t} \pb{u_t,x_t|q_{[t-1]}} \pb{q_t|u_t,x_t,q_{[t-1]}}} \nonumber\\
	& \utag{a}{=} \frac{ \sum_{x_t} \pb{u_t,x_t|q_{[t-1]}}  \wb{q_t|u_t,x_t}}{\sum_{x_t,u_t} \pb{u_t,x_t|q_{[t-1]}} \wb{q_t|u_t,x_t}} \nonumber\\
	& \utag{b}{=} \frac{ \sum_{x_t} \pb{u_t,x_t}  \wb{q_t|u_t,x_t}}{\sum_{x_t,u_t} \pb{u_t,x_t} \wb{q_t|u_t,x_t}}, \label{eq:ut-qt}
\end{align}
where \uref{a} follows because $Q_t$ is a stochastic function of $\{U_t,X_t\}$ given in \eqref{eq:encoding-function}, and \uref{b} follows because $\{u_t,x_t\} \subseteq \cB_{t-1}$ and the inducative assumption \eqref{eq:inductive}.


From \eqref{eq:encoding-function}, we have
\begin{align*}
& \pb{u_t,x_t}  \wb{q_t|u_t,x_t} \hspace{-1pt} =	\hspace{-3pt}\begin{cases}
	\pb{u_t}\pi(x_t), & \hspace{-2pt} |q_t|=1, \\
	\pb{u_t,x_t}- \pb{u_t}\pi(x_t), & \hspace{-2pt} |q_t| = 2.
	\end{cases}
\end{align*}

For $|q_t|=1$, \eqref{eq:ut-qt} can be written as 
\begin{align}
	\pb{u_t|q_{[t]}} 
	& = \frac{ \sum_{x_t} \pb{u_t,x_t}  \wb{q_t|u_t,x_t}}{\sum_{x_t,u_t} \pb{u_t,x_t} \wb{q_t|u_t,x_t}} \nonumber \\
	& = \frac{ \sum_{x_t} \pb{u_t}\pi(x_t)}{\sum_{x_t,u_t} \pb{u_t}\pi(x_t)} \nonumber \\
	& \utag{a}{=} \frac{  \pb{u_t} \sum_{x_t}\pi(x_t)}{\sum_{x_t}\pi(x_t) \sum_{u_t} \pb{u_t} } \nonumber \\
	& = \pb{u_t}, \label{eq:temple1}
\end{align}
where \uref{a} follows because $\pi(x_t)$ is independent of $u_t$.

Similarly, for $|q_t|=2$, \eqref{eq:ut-qt} can be written as 
\begin{align}
	\pb{u_t|q_{[t]}} 
	& = \frac{ \sum_{x_t} \pb{u_t,x_t}  \wb{q_t|u_t,x_t}}{\sum_{x_t,u_t} \pb{u_t,x_t} \wb{q_t|u_t,x_t}} \nonumber \\
	& = \frac{ \sum_{x_t} \pb{u_t,x_t}- \sum_{x_t}\pb{u_t}\pi(x_t)}{\sum_{x_t,u_t} \pb{u_t,x_t}- \sum_{x_t,u_t}\pb{u_t}\pi(x_t)} \nonumber \\
	& = \frac{\pb{u_t}- \pb{u_t}\sum_{x_t}\pi(x_t)}{1- \sum_{x_t}\pi(x_t)} \nonumber \\
	& = \pb{u_t}. \label{eq:temple2}
\end{align}

From \eqref{eq:temple1} and \eqref{eq:temple2}, we can obtain that 
\begin{equation}
\label{eq:temple3}
    I\left(U_t;Q_{[t]}\right)=0.
\end{equation}
Therefore, by plugging \eqref{eq:temple3} into \eqref{eq:last-step} and using  Proposition~\ref{proposition:1}, we obtain  
\[I\left(X_{\cB_{t}};Q_{[t]}\right)=0,\]
which concludes our induction proof.



 
\subsection{Proof of  Proposition~\ref{proposition:1}}
First, we have
\[ I_1 = I\left(X_{\cB_{t}};Q_{[t-1]}\right) 
    \utag{a}{\leq} I\left(X_{\cB_{t-1}};Q_{[t-1]}\right) \utag{b}{=}  0,\]
where \uref{a} follows because $\cB_{t-1} =\cB_{t}\cup\{t\}$ by definition, and \uref{b} follows from the inductive assumption \eqref{eq:inductive}. Second,
\[I_2 = I\left(U_t;Q_{[t-1]}\right) \utag{a}{\leq} I\left(X_{\cB_{t}};Q_{[t-1]}\right) = I_1 \leq 0,\]
where \uref{a} follows because $U_t \subseteq X_{\cB_{t}}$ by definition.

Finally, we prove  that $I_3=0$ as follows
\begin{align}
	I_3 & =  I\left(X_{\cB_{t}}\setminus U_t;Q_t|U_t,Q_{[t-1]}\right) \nonumber\\
	& ~~~ \utag{a}{\leq} I\left(X_{\cB_{t}}\setminus U_t;U_t,X_t,S_t|U_t,Q_{[t-1]}\right) \nonumber\\
	& ~~~ \utag{b}{=} I\left(X_{\cB_{t}}\backslash U_t;X_t|U_t,Q_{[t-1]}\right) \nonumber\\
	& ~~~ = I\left(X_{\cB_{t}}\setminus U_t;X_t|U_t\right) +  I\left(X_{\cB_{t}}\setminus U_t;Q_{[t-1]}| X_t,U_t\right) \nonumber\\
	&~~~~~~ -  I\left(X_{\cB_{t}}\setminus U_t;Q_{[t-1]}| U_t\right) \nonumber\\
	& ~~~ \utag{c}{=} I\left(X_{\cB_{t}}\setminus U_t;X_t|U_t\right), \label{eq:second-term}
\end{align}
where \uref{a} follows because $Q_t$ is encoded from $\{U_t,X_t,S_t\}$, \uref{b} follows because $S_t$ is independent of $\{X_i:i \in \mathbb{N}\}$ and $Q_{[t-1]}$, and \uref{c} can be justified 
because one can check that
\[ I\left(X_{\cB_{t}}\backslash U_t;Q_{[t-1]}| X_t,U_t\right) =   I\left(X_{\cB_{t}}\backslash U_t;Q_{[t-1]}| U_t\right) = 0\]
from $\cB_{t-1} =\cB_{t}\cup\{t\}$ and the inductive assumption \eqref{eq:inductive}. 

To finish proving $I_3 = 0$, we claim that 
\[I\left(X_{\cB_{t}}\setminus U_t;X_t|U_t\right) =0.\]
Towards this end, by letting $\cB^{-}_{t} = \{i:i \leq t, F_i=\text{ON}\}\setminus\{F^{-}(t)\}$,
and $ \cB^{+}_{t} = \{i:i \geq t+2\}$, we can easily obtain from the Markovity of $\{X_i:i \in \mathbb{N}\}$ that
\begin{align*}
	I\left(X_{\cB_{t}}\backslash U_t;X_t|U_t\right) 
	& = I\left(X_{\cB^{-}_{t}},X_{\cB^{+}_{t}} ; X_t|X_{\{F^{-}(t),t+1\}} \right)  = 0,
\end{align*}
 which concludes that $I_3 = 0$.



 

\section{Proof of Theorem~\ref{thm:main}: Converse}
\label{sec:converse}

To obtain an upper bound on the rate $R_t$, we derive a lower bound  on the average downloading cost $\mathbb{E}\left[|Q_t|\right]$, which can be 
obtained by solving the following optimization problem:
\begin{equation}
\label{eq:Optimazation}
\begin{aligned}
& \underset{\pb{u_t,x_t,q_t}}{\text{minimize}}
& & \mathbb{E}\left[|Q_t|\right] = \sum_{q_t} \pb{q_t}|q_t| &\\
& \text{subject to}
& & \pb{x_t,q_t} =0, \ x_t \notin q_t,  & \text{(decodability)}\\
& & & \pb{q_t|u_t} = \pb{q_t}. & \text{(relaxed privacy)} 
\end{aligned}
\end{equation}
Here, the relaxed privacy constraint is obtained by relaxing our original  privacy requirement  $I\left(Q_{[t]};X_{\cB_{t}}\right) = 0$ to  $I\left(Q_{t};U_t\right) = 0$. This is a relaxation because  $\{F^{-}(t),t+1\} \subseteq \cB_t$.

For clarity, we illustrate all feasible $\pb{u_t,x_t,q_t}$ in Table~\ref{table:marginal} with two auxiliary variables $z_1$ and $z_2$, where 
\[z_1 = \Pb{Q_t=A|U_t=(A,A)},\]
and
\[z_2 = \Pb{Q_t=B|U_t=(A,A)}.\]
Clearly, all entries in Table~\ref{table:marginal} must be non-negative.

Then the optimization problem given in \eqref{eq:Optimazation} can be re-written as
\begin{equation}
\label{eq:Optimazation-1}
\begin{aligned}
& \underset{z_1,z_2}{\text{minimize}}
& & \mathbb{E}\left[|Q_t|\right]= 2- z_1  - z_2 \\
& \text{subject to}
& & 0 \leq z_1 \leq \pi(A) , \\
& & & 0 \leq z_2 \leq \pi(B),
\end{aligned}
\end{equation}
where $\pi(A)$ and $\pi(B)$ are defined in \eqref{eq:pi}.

We can easily see that the optimal value to the problem in \eqref{eq:Optimazation-1} is given by 
\[\min_{z_1,z_2} (2- z_1  - z_2) = 2- \pi(A) - \pi(B),\]
which completes the proof that
\[\frac{1}{R_t} \geq 2- \sum_{x_t \in \{A,B\}} \pi(x_t).\]



\begin{table}[t!]	
      \centering
    \begin{tabular}{ c c |c c c }
	$U_{t}$ & $X_t$& $Q_t=A$ &$Q_t=B$& $Q_t=\{A,B\}$\\
	 \toprule
	 $(A,A)$ & $A$ & $ z_1 p_{aa} $ &$0$   &$p_{aa} \left(p_{a|aa} - z_1 \right) $\\
	 $(A,A)$ & $B$ & $0$ 	&$z_2 p_{aa} $ &$p_{aa}  \left(p_{b|aa} - z_2 \right)$\\
	 $(A,B)$ & $A$ & $ z_1 p_{ab}$	&$0$	&$p_{ab} \left(p_{a|ab} - z_1 \right)$\\
	 $(A,B)$ & $B$ & $0$	&$ z_2 p_{ab}$	&$p_{ab}  \left(p_{b|ab} - z_2 \right)$\\
	 $(B,A)$ & $A$ & $ z_1 p_{ba}$ &$0$	&$p_{ba}  \left(p_{a|ba} - z_1 \right)$\\
	 $(B,A)$ & $B$ & $0$ 	&$ z_2 p_{ba}$ &$p_{ba} \left(p_{b|ba} - z_2 \right)$\\
	 $(B,B)$ & $A$ & $ z_1 p_{bb}$	&$0$	&$p_{bb} \left(p_{a|bb} - z_1 \right)$\\
	 $(B,B)$ & $B$ & $0$	&$ z_2 p_{bb}$	&$p_{bb} \left(p_{b|bb} - z_2 \right)$\\
	\end{tabular}
	  \caption{The joint distribution $\pb{u_t,x_t,q_t}$ satisfying the decodability and the privacy constraint, where $p_{aa}$ denotes $\Pb{U_t=(A,A)}$, and $p_{a|aa}$ denotes $\Pb{X_t=A|U_t=(A,A)}$. Both are constants given by the transition matrix of the Markov chain. 
	  $z_1 = \Pb{Q_t=A|U_t=(A,A)}$ and $z_2=\Pb{Q_t=B|U_t=(A,A)}$ are two variables in the optimization problem.
	  }
	   \label{table:marginal}
\end{table}

\section{Conclusion and discussion}
In this paper, we study the problem of preserving ON-OFF privacy for past and future requests with a two-states Markov source. The ON-OFF privacy problem \cite{Naim_2019} was introduced to capture the scenario that the privacy may be switched between ON and OFF. Different from the setup in \cite{Naim_2019}, wherein only the past requests for which privacy turned ON are preserved, a more stringent privacy requirement is studied. In this paper, we construct ON-OFF privacy scheme for $N=2$ sources and prove their optimality. Both achievability and converse proof given in this paper can be easily generalized to more than two sources.

\end{document}